\documentclass[prd,aps,12pt,tightenlines]{revtex4}
\def\be{\begin{equation}}
\def\ee{\end{equation}}
\def\ba{\begin{eqnarray}}
\def\ea{\end{eqnarray}}
\def\half{{1 \over 2}}
\def\third{{1 \over 3}}
\def\g{g_{YM}}
\def\Etot{E_{\rm tot}}
\def\Itot{I_{\rm tot}}
\def\Ftot{F_{\rm tot}}
\def\ene{{\cal E}}

\usepackage{graphicx}

\begin{document}
\title{Classical thermodynamics of gravitational collapse}
\author{Z.~Gecse and S.~Khlebnikov}
\affiliation{
Department of Physics, Purdue University, West Lafayette, IN 47907,
USA}
\begin{abstract}
We study numerically gravitational collapse of a spherically symmetric instanton 
particle in five dimensions. We show that the late stages of the process are 
characterized by a nearly constant ``free energy'', the value of
which matches (within numerical uncertainties) the value obtained from standard
black-hole thermodynamics. This suggests a purely classical interpretation of the
free energy of a black hole.
\end{abstract}
\maketitle

\section{Introduction}
Brane-world scenarios \cite{Joseph,Rubakov:1983bb} provide a new motivation for 
studying energy relations associated with gravitational collapse. In these scenarios,
instanton transitions \cite{Belavin:1975fg,'tHooft:1976up} on the brane can be viewed 
as transport of the topological charge through the brane, much like a change of magnetic 
flux through a superconducting ring can be viewed as transport of vortices through
the ring. In this picture, a $\theta$-vacuum \cite{Callan:1976je,Jackiw:1976pf} 
corresponds to a steady flow of the topological charge into extra dimensions, 
and preventing such a flow amounts to a solution to the strong CP problem. 
The simplest mechanism that can achieve that is an energy cost 
\cite{Khlebnikov:2004am,Khlebnikov:2006yq}. Indeed, while
in the 4-dimensional chromodynamics instantons connect states that are exactly 
degenerate in energy (a consequence of the symmetry with respect to the ``large'' gauge
transformations \cite{Callan:1976je,Jackiw:1976pf}), in the presence of extra dimensions 
that need not be so \cite{Khlebnikov:1987zg}. 

In one version of the scenario, energy accumulates in an extra-dimensional black 
hole \cite{Khlebnikov:2007ii}.
However, it is not clear from the outset what is the appropriate definition of energy
in this case. Should we be talking about the usual mass of a black hole, 
as defined by the spatial asymptotics of the metric, or is there a different
``free energy'' associated with late stages of gravitational collapse? 
Evolution of a quantum state on the brane is determined by $\exp[iI(t)]$, where 
$I$ is the action of matter and gravity (including those in the bulk). Based
on this, in the present work, we explore properties of the following quantity
\be
F(t)  = - \frac{dI(t)}{dt} \; ,
\label{F}
\ee
i.e., minus the Lagrangian computed on the collapsing solution.
In a brane-world scenario, there is a preferred time coordinate---the proper time
of an observer on the brane, and it is this time that we use in the definition (\ref{F}).

We consider evolution of an instanton
``particle'' in the 5-dimensional (5d) $SU(2)$ Yang-Mills (YM) theory coupled to gravity
in asymptotically flat spacetime. Although this asymptotically flat case is not
a realistic braneworld (gravity and the gauge field are not localized on the brane), 
the question about time-dependence of $F$ can still be asked; by a ``brane''
in this case we mean a sphere at large radius.
Note that the derivative in (\ref{F}) is a full derivative, rather than a partial
one taken at fixed endpoint values of the canonical coordinates; the partial derivative 
would give us the mass $E$ of the black hole.

We refer to $F$ as the free energy (of a classical black hole), for the following 
reason. Suppose at late times
$F(t)$ becomes a constant, i.e., the action becomes a linear function of time.
Such a function can be analytically continued to the Euclidean time $\tau = -it$,
so that the Euclidean action is $I_E = F\tau$.
Since we are dealing with just a single classical trajectory, the partition sum
reduces to $\exp(-F/T)$ (where $T$ is the Hawking temperature \cite{Hawking:1974sw}), 
which means that $F$ is the free energy.

Interpretation of the Euclidean Lagrangian as a thermodynamic potential
is known in the literature \cite{Gibbons:1976ue},
and, as we will see, our result agrees with the corresponding thermal expression,
Eq.~(\ref{thermo}) below. However, this agreement is by no means obvious a priori: 
the thermal result
is obtained by considering the stationary vacuum exterior alone, while our 
result corresponds to a time-dependent metric of collapsing matter. Moreover,
during collapse, $F(t)$ is expected to change, and a priori it is not even
clear if it will approach {\em any} constant value at late times.

The latter point, however, can be checked directly
in the case of spherical symmetry, when collapse can be simulated numerically.
In this paper, we present results of such a simulation.

Gravitational instability of an initially static instanton particle in the 5d pure 
YM theory follows directly from the classical scale invariance of the corresponding
4d theory and had been discussed previously \cite{Volkov:2001tb}.
Here, we discuss the process of gravitational collapse of such objects 
(we have considered both static and nonstatic initial configurations).
We observe formation of a black hole (of some energy $E$) and find that soon after
that $F$ reaches a nearly constant value $F\approx \third E$.

We interpret the factor $\third$ as follows. For a 5d Schwarzschild black hole,
the Hawking temperature, computed by the method of Ref.~\cite{Gibbons:1976ue}, 
is $T = \hbar/4\pi r_0$, where $r_0$ is the gravitational radius:
\be
r_0^2= \frac{2}{3\pi} G_5 E \; ,
\label{r_0}
\ee
and $G_5$ is 5d Newton's constant (our choice of the radial coordinate $r$ will
become apparent shortly). The entropy \cite{Bekenstein:1973ur,Hawking:1974sw} is
$S = 4\pi^2 r_0^3/\hbar G_5$. Then, by standard thermodynamics, the free energy
is
\be
F_{\rm thermo} = E - ST = \third E \; .
\label{thermo}
\ee
Note that, while $T$ and $S$ each contain a power of the Planck constant $\hbar$, 
the product $ST$ does not and so has a finite classical limit. We see that the
value (\ref{thermo}) agrees with the value of $F$
reached in the course of our classical simulation.

For the Schwarzschild 
exterior alone, Eq.~(\ref{F}) would give $F=E$. We find that the difference $F-E$ comes 
from a thin shell near the horizon. It may seem, then, 
that $F$ being nearly constant
at large times is simply a consequence of the slowing down (redshift) of 
the near-horizon dynamics as seen by a distant observer. If the near-horizon evolution
were to
stop entirely, however, we would again have $F=E$. (This relation holds for any static 
configuration, vacuum or not). This not being the case means that
the fields continue to change, and their time dependence significantly contributes
to $F$. The situation is similar to ordinary statistical 
equilibrium: macroscopically nothing changes, but there is still motion on 
the microscopic scale.

In the remainder of the paper, we describe various technical details of our 
approach. We conclude with some comments on the quantum behavior of the system.

\section{Equations of motion in spherical symmetry}  

\subsection{Spherically-symmetric ansatz}
We use the isotropic coordinates, in which a spherically-symmetric metric takes 
the form
\be
\label{isoMetric}
ds^2 = -N^2(t,r) d t^2 + \Psi^2(t,r)(d r^2 + r^2 d \Omega_3^2) \; .
\ee
These are convenient for numerical studies of gravitational collapse because
of the absence of singularities in the metric functions: 
formation of a black hole 
is reflected in vanishing of the lapse function $N$ at some point, while 
the conformal factor $\Psi$
remains finite everywhere. This observation was used previously in 
Ref.~\cite{Finelli:2000gi}
for studying the evolution of metric perturbations after inflation.

The 5d Schwarzschild geometry in the isotropic coordinates has the form
\be
\label{SchwarzIso}
ds^2=
-\frac{\left(1-\frac{r_0^2}{r^2}\right)^2}{\left(1+\frac{r_0^2}{r^2}\right)^2} d t^2 
+ \left( 1+\frac{r_0^2}{r^2} \right)^{2}
\left(d r^2 + r^2 d \Omega_3^2 \right) \; ,
\ee
where $r_0$ is the gravitational radius. Note that $\Psi$ does not stay finite
in the ``white-hole'' region of this geometry---there is a singularity at $r=0$.
We find, however, that, while the exterior metric of black holes that we obtain
numerically well matches the $r > r_0$ region of the metric (\ref{SchwarzIso}), the 
interiors are quite different, and in our case $\Psi$ stays finite.

When the metric is asymptotically flat, the total energy 
(ADM mass \cite{Arnowitt:1962hi}) of the spacetime is
\be
\Etot = -\frac{3\pi}{4 G_5} R^3 \left. \partial_r \Psi \right|_{r=R} \; ,
\label{Etot}
\ee
where $R$ is some large radius. In the limit
$R\to \infty$, $\Etot$ will be conserved. 
One should keep in mind, though, that, for a general initial 
state, only part of the mass will collapse to form a black hole, while the rest 
will be radiated away in the form of matter waves. 
The expression (\ref{Etot}) is the total
and does not, in general, coincide with the mass $E$ of the black hole.

Related to the isotropic coordinates are Cartesian coordinates
$x^i$, $i=1,\ldots,4$, such that
$d r^2 + r^2 d \Omega_3^2 = (dx^i)^2$. In these Cartesian coordinates,
our ansatz for the $SU(2)$ gauge field is
\be
\label{gaugeAnsatz}
A_\mu^a = \left( 0, \eta^a_{ij}n_j \frac{f(t,r)}{r} \right) \; ,
\ee
where $\mu= 0,\ldots,4$; $a=1,2,3$;
$n_j$ is the unit radial vector, $\eta^a_{ij}$ are the t'Hooft symbols 
\cite{'tHooft:1976up}, and $f(t,r)$ is the dynamical variable.
Note that $A_0^a = 0$ at all times, which is consistent with the equations of motion.

For a gauge field of the form (\ref{gaugeAnsatz}), the topological charge
\be 
Q=-\frac{1}{64\pi^2}
\int d^4 x \epsilon^{ijkl}F^{a}_{\ ij}F^a_{\ kl}
\ee
($F^a_{\ \mu\nu}$ is the field strength) becomes
\be
Q=-\frac{3}{4} \int f(f-2)f' dr = - \frac{1}{4} \int \frac{d}{dr} ( f^3 - 3f^2) dr \; .
\ee
So, any configuration that interpolates between $f=0$ at $r=0$ and $f=2$ at large $r$
has unit topological charge. An example is the BPST instanton \cite{Belavin:1975fg},
\be
\label{selfDual}
f_0(r) = \frac{2r^2}{\lambda^2+r^2} \; ,
\ee
where $\lambda$ is an arbitrary scale parameter.

\subsection{Equations of motion}
The part of the Einstein-YM equations that are evolution equations
can be written in the Hamiltonian form
\ba
\label{EOM1}
\frac{\dot{f}}{N}&=&\frac{p}{\Psi^2},\\
\label{EOM2}
\frac{\dot{p}}{N}&=&f''+\frac{f'}{r}+\frac{N'}{N}f' - \frac{\partial V}{\partial f},\\
\label{EOM3}
\frac{\dot{\Psi}}{N}&=&-\frac{K\Psi}{4},\\
\frac{\dot{K}}{N}&=&\frac{K^2}{2} +\frac43 \kappa^2 \ene
- 8\frac{\kappa^2}{\g^2\Psi^4r^2}V
-4\frac{\Psi'}{\Psi^3}\left(\frac{\Psi'}{\Psi}+\frac2r \right)
-4\frac{N'}{N\Psi^2}\left(\frac{\Psi'}{\Psi}+\frac1r \right).
\label{EOM4}
\ea
Here $\kappa^2 = 8 \pi G_5$, dots and primes denote temporal and spatial derivatives, 
respectively,
\be
V=\frac{f^2(f-2)^2}{2r^2} \; ,
\ee
and
\be
\label{gaugeEnergy}
\ene
=\frac{3}{\g^2\Psi^4r^2}\left(\frac12 \frac{p^2}{\Psi^2}+\frac12 f'^2 + V\right) \; .
\ee
$V$ can be thought of as a potential for the gauge field, and $\ene$ as the field's
energy density.
The instanton (\ref{selfDual}) is then a ``kink'', which interpolates between two
degenerate minima of $V$, $f=0$ and $f=2$.

We see that $K$ (the trace of the extrinsic curvature) and $p$ can be regarded 
as conjugate momenta of $\Psi$ and $f$.
The lapse function $N$, on the other hand, has zero conjugate momentum, i.e., it is
a Lagrange multiplier. The Einstein equation for it is an ordinary
differential equation that has to be solved at each moment of time
\be
\label{eqN}
\frac{r}{N}\left(\frac{N'}{r\Psi^2}\right)'
+ 2\frac{r}{\Psi}\left(\frac{\Psi'}{r\Psi^2}\right)' = 
\frac{-\kappa^2}{\g^2\Psi^4r^2}\left( \frac{p^2}{\Psi^2} + 2 f'^2 - 4V \right).
\ee
The remaining Einstein equations are constraints: the energy constraint
\be
\label{energyConstr}
-3\frac{\nabla^2\Psi}{\Psi^3} = \kappa^2 \ene - \frac{3}{8} K^2 \; ,
\ee
where $\nabla^2$ is the flat-space Laplacian, and the momentum constraint 
\be
\label{momentumConstr}
\frac14 K' = \frac{\kappa^2}{\g^2\Psi^4r^2}pf' \; .
\ee
If these are satisfied at the initial time slice, the evolution equations together
with Eq.~(\ref{eqN}) ensure that they are satisfied at all times.

Eqs.~(\ref{EOM2}) and (\ref{EOM4}) require boundary conditions
at $r=R$. We choose 
\be
\label{bcR}
K'(t, R) = p'(t, R) = 0 \; .
\ee
Although this does not automatically enforce the momentum constraint 
(\ref{momentumConstr}) at $r=R$, for a large enough $R$ the constraint holds
there to a sufficient accuracy.

Regularity of the gauge field at $r=0$ implies that $f(t,r)=O(r^2)$ and 
$p(t,r)=O(r^2)$ at all times. Using these in the momentum constraint 
(\ref{momentumConstr}), we find
\be
\label{bc0}
K'(t, 0) = f(t, 0) = 0 \; .
\ee
Together with Eq.~(\ref{bcR}), this provides enough boundary conditions for 
the evolution equations.

We still need boundary conditions for Eq.~(\ref{eqN}).
The metric ansatz (\ref{isoMetric}) leaves us the freedom of a global ($r$-independent)
redefinition of time. We use this freedom to set $N(t, R) = 1$. 
If the boundary is far away from any matter, $N$ at large $r$
is close to its Newtonian limit, for which
\be
\label{newt}
N'(t, R) = - \frac{2 \Psi'(t, R)}{\Psi(t,R)} \; .
\ee
For large enough $R$, the difference of this from zero is inessential, and 
the data presented here have been obtained with
\be
\label{bcN}
N'(t, R) = 0 \; .
\ee

Finally, we list a convenient integrated form of the energy constraint 
\be
\Etot = 2\pi^2 \int_0^R \left (\ene - \frac{3}{8\kappa^2} K^2 \right) \Psi^3 r^3 dr \; .
\label{eneInt}
\ee
The constraints are not involved in the evolution algorithm but are used to monitor
the accuracy of the simulation.

\subsection{The action}
The original action of the model is
\be
\Itot =\int d^5 x \sqrt{-g^{(5)}}\left( \frac{R^{(5)}}{16\pi G_5} 
-\frac{1}{4\g^2} F^a_{\ \mu\nu}F^{a\mu\nu}\right) + I_B \; ,
\label{Iorig}
\ee
where $I_B$ is the boundary term chosen so as to remove the second derivatives of the
metric functions \cite{LL}. Just as we did for the energy, we need to distinguish between
the total action (\ref{Iorig}) and the part of it, which we call $I$, that
corresponds to the black hole alone. This is possible at sufficiently late times,
when the black hole and the outgoing radiation become well separated.

When computed on our spherically-symmetric ansatz, the total action 
becomes
\be
\Itot = \int dt \int_0^R dr L \; ,
\label{Itot}
\ee
where
\ba
&L=&\frac{2\pi^2}{16\pi G_5} \left(-12\frac{\dot{\Psi}^2\Psi^2r^3}{N} 
+6\Psi'^2Nr^3 +6\Psi'N'\Psi r^3\right) \nonumber\\
&&-\frac{2\pi^2}{4\g^2} N r\left(6f'^2 -6\frac{\dot{f}^2\Psi^2}{N^2} 
+12V\right) \; .
\label{L}
\ea
A {\em static} configuration is one for which $\dot{\Psi}$ and $\dot{f}$ at a given
instant vanish. For any such
configuration, one can integrate by parts the term with $N'$ in (\ref{Itot}) and use 
the energy constraint (\ref{energyConstr}) to bring the
Lagrangian to the form
\be
\frac{d\Itot}{dt}
= \frac{3\pi}{4 G_5} R^3 \left. N \Psi \Psi' \right|_{r=R} \; .
\label{Lstatic}
\ee
In the asymptotically flat case, $N$ and $\Psi$ approach unity at large $r$, and
the right hand side of Eq.~(\ref{Lstatic}) becomes minus the energy (\ref{Etot}). 
This is an expected 
result. Indeed, for static configurations the total and partial derivatives of 
the action coincide, so the Lagrangian equals minus the Hamiltonian.

\subsection{Initial state}
\label{subsect:init}
In general, an initial state is an arbitrary solution of the constraints
(\ref{energyConstr}) and (\ref{momentumConstr}). One class
of such states are static states,
for which the momenta vanish identically, so the momentum constraint is
trivially satisfied. To solve the energy constraint for such states, 
we pick a function $f(r)$ for which 
gravity is weak and solve the constraint equation
\be
- \Psi\nabla^2 \Psi =  \frac{\zeta}{r^2}
\left(  \half f'^2 + V \right) 
\label{ec_static}
\ee
by a formal expansion in the parameter $\zeta \equiv \kappa^2 / g_{YM}^2$.
The actual expansion parameter is $\zeta / \lambda^2$, where $\lambda$ is the
spatial size of the configuration.
For example, if $f(r)$ is taken to be the instanton (\ref{selfDual}), 
the first three terms in the expansion are
\be
\Psi = 1 + \frac{2 \zeta (2\lambda^2+r^2)}{3 (\lambda^2+r^2)^2}
-\frac{2 \zeta^2 (27\lambda^6 + 48\lambda^4r^2 + 32\lambda^2r^4 + 8r^6)}
        {45 \lambda^2(\lambda^2+r^2)^4}+\dots \; .
\label{Psi}
\ee
Retaining a sufficient number of terms allows us to solve the constraint
to any required accuracy. 

Using a large static instanton as the initial condition offers certain advantages
for numerical work. These are related to the fact that such an instanton is
nearly a classical solution (it is destabilized only
by gravity, which for large $\lambda$ is weak). We find that in this case collapse
generates relatively little of an outgoing wave, and the total action $\Itot$ 
nearly coincides with the action
of the black hole, $I \approx \Itot$. This simplifies
the measurement of $I$. [In general, $I$ has to be obtained by integrating over $r$ 
in (\ref{Itot}) not from $0$ to $R$ but from $0$ to some intermediate radius, 
so that the outgoing wave is left out.]

Initial stages of the collapse of an instanton with a large $\lambda$ can be
studied analytically, since
at large $\lambda$ the system is in the Newtonian limit. The equations of motion
(\ref{EOM1})--(\ref{EOM2}) in this limit combine into
\be
\ddot{f} = f'' + \frac{f'}{r} + \frac{N'}{N} f' - \frac{\partial V}{\partial f} \; .
\label{eqf}
\ee
We search for a solution in the form
\be
f(t,r) = \frac{2r^2}{r^2 + \lambda^2(t)} + \dot{\lambda}^2 h_1(r,\lambda) 
+ \ddot{\lambda} h_2(r,\lambda) \; .
\label{solf}
\ee
This describes the slow motion of the instanton size (``modulus'')  $\lambda$,
together with small deformations of the instanton shape that this motion causes.

Now, in terms of the parameter $\zeta$ introduced above, $\dot{\lambda}$ is $O(\sqrt{\zeta})$,
while $\ddot{\lambda}$ is $O(\zeta)$. So, in (\ref{eqf}) we need $N'/N$ only to order
$\zeta$. Using Eqs.~(\ref{eqN}) and (\ref{Psi}), we find
\be
\frac{N'}{N} = - \frac{2\Psi'}{\Psi} + O(\zeta^2) = 
\frac{8\zeta}{3} \left( \frac{r}{(\lambda^2 + r^2)^2} 
+ \frac{2r\lambda^2}{(\lambda^2 + r^2)^3} \right) + O(\zeta^2) \; .
\label{N'}
\ee
Substituting Eqs.~(\ref{solf}) and (\ref{N'}) into Eq.~(\ref{eqf}) and projecting
onto the instanton zero mode $\psi_0 = \partial f_0 / \partial \lambda$,
we obtain an equation for $\lambda(t)$,
\be
- \ddot{\lambda} = \frac{8\zeta}{15\lambda^3} \; .
\label{eql}
\ee
The solution that has a maximum size $\lambda_0$ at time $t= 0$ is
\be
\lambda(t) = \left( \lambda_0^2 - \frac{8\zeta t^2}{15\lambda_0^2}  \right)^{1/2} \; .
\label{soll}
\ee
This allows one, for instance, to determine the time it takes the instanton to collapse
(see below).

We have also considered nonstatic initial conditions with $f$ 
of the form (\ref{selfDual}) and the momentum $p$ of the form
\be
p(t, r) = \frac{2 v r^2}{(\lambda^2 + r^2)^2}  \; ,
\ee
where $v$ is a parameter. 
(This does not quite satisfy the boundary condition (\ref{bcR}), but for large
$R$ the mismatch is small.)
The results are similar to those presented below
for the case $v = 0$.

\section{Numerical results}
\label{sec:results}
Following Ref.~\cite{Finelli:2000gi}, to improve stability of the numerical method
we add to the right-hand side 
of (\ref{EOM4}) a term 
proportional to the energy constraint. (This brings in the flat-space Laplacian  
of $\Psi$.)
We choose the unit of length so that the gravitational radius (\ref{r_0}) is
$r_0 = 1$ for $E = E_{\rm inst}=8\pi^2 / g_{YM}^2$ 
(the instanton mass in the absence of gravity).
This is equivalent to setting $\zeta=\kappa^2 /\g^2=\frac{3}{2}$ for the only 
parameter that appears in the equations of motion. The energy and the Lagrangian are
made dimensionless by measuring them in units of $E_{\rm inst}$. 
In these units, Eq.~(\ref{r_0}) becomes
\be
E = r_0^2 \; \; .
\label{eneBH}
\ee
A change of the spatial coordinate
\be
\label{changeCoord}
r \to x = \frac{r}{r+1}
\ee
allows us to discretize the equations on a uniform grid (in $x$), while maintaining
a higher density of grid points (in $r$) in the region of main interest $r\sim 1$.
Results are presented, however, in terms of the original coordinate $r$.

\begin{figure}[ht]
\includegraphics[scale=.75]{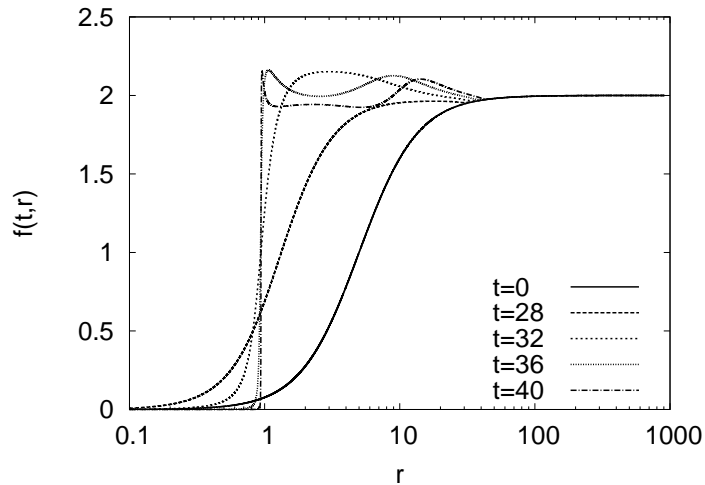}
\caption{$f(t,r)$ as a function of $r$ at different moments of time.}
\label{fig:f}
\end{figure}

\begin{figure}[ht]
\includegraphics[scale=.75]{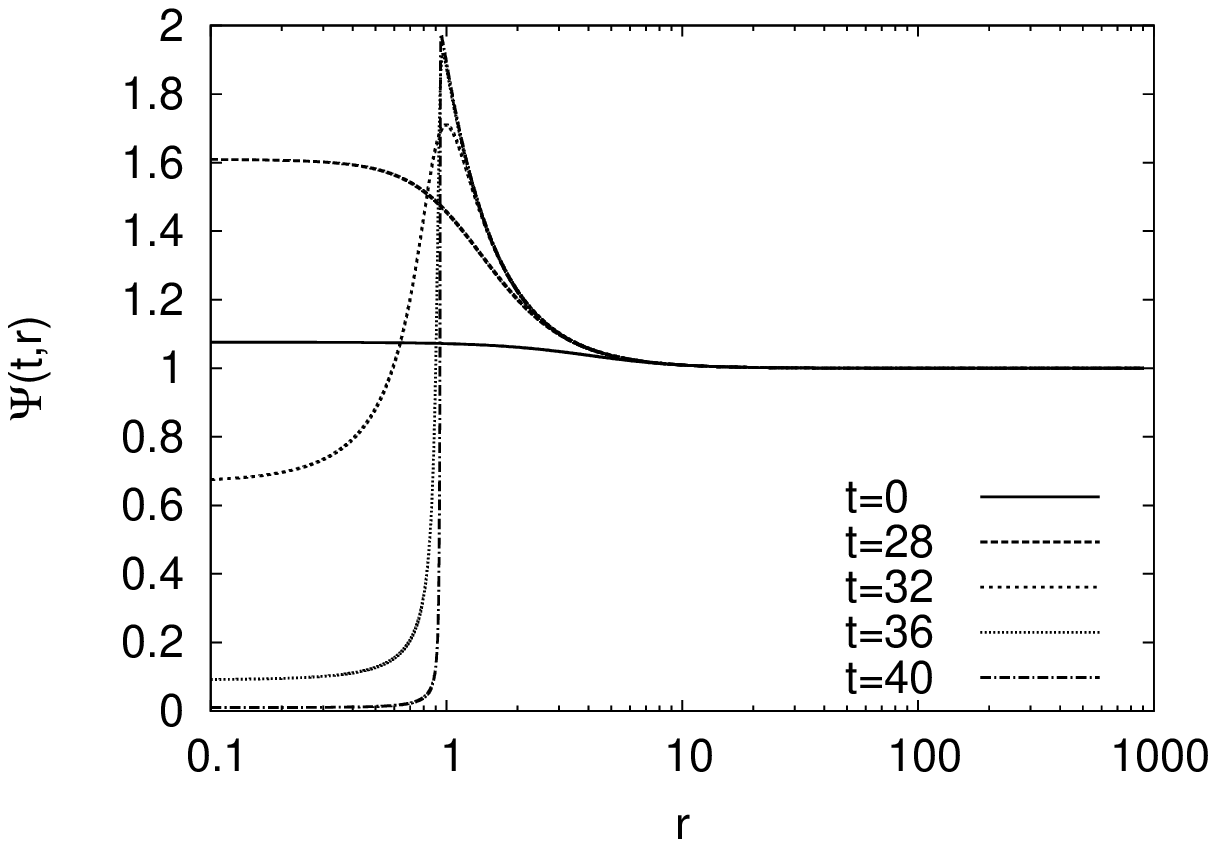}
\caption{$\Psi(t,r)$ as a function of $r$ at different moments of time.}
\label{fig:s}
\end{figure}

\begin{figure}[ht]
\includegraphics[scale=.75]{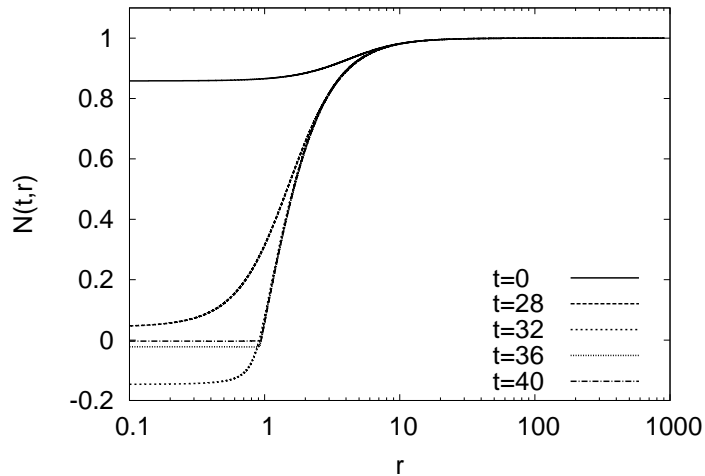}
\caption{$N(t,r)$ as a function of $r$ at different moments of time.}
\label{fig:l}
\end{figure}

These results are for the static initial state (\ref{selfDual}) with $\lambda=5$.
Figures~\ref{fig:f}--\ref{fig:l} 
show the time evolution of the profiles of the three dynamical
functions, $f$, $\Psi$, and $N$. Formation of a black hole is signaled by 
$N(r,t)$ (as a function of $r$) crossing zero. In the present case, this happens at 
$t\approx 28$. Note that the rapid evolution shown in the figures follows a prolonged
stage during which little happens. The duration of that stage depends on the initial
instanton size and, for large $\lambda$, 
can be estimated using Newtonian gravity: using $\zeta = \frac{3}{2}$
in Eq.~(\ref{soll}), we find
\be
t = \frac{\sqrt{5}}{2} \lambda^2 \; ,
\label{t}
\ee
which is well born out by our numerical results.

At late stages, the metric functions outside the horizon approach the Schwarzschild
form (\ref{SchwarzIso}), while inside the horizon they approach a curious collapsing
universe with nearly space-independent $N$ and $\Psi$.

\begin{figure}[ht]
\includegraphics[scale=.75]{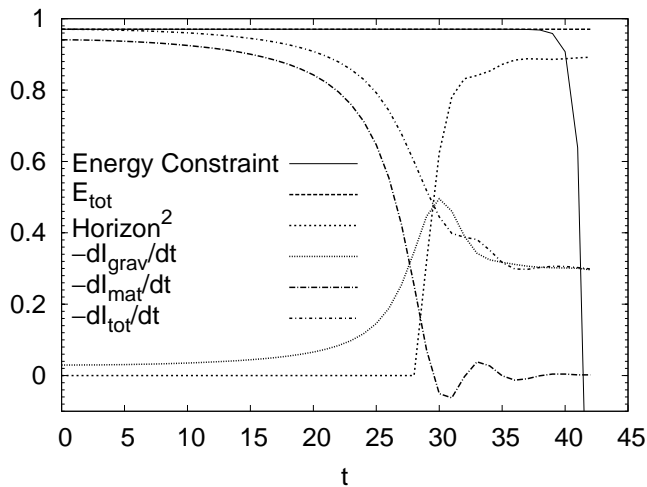}
\caption{Dependence of various quantities on time. $dI_{\rm tot}/dt$ is
the sum of the gravity ($dI_{\rm grav}/dt$) and matter ($dI_{\rm mat}/dt)$
contributions, each of which is the integral over $r$ of the corresponding term
in Eq.~(\ref{L}). ``Horizon$^2$'' is the square of the radius at which
$N(t,r)$ crosses zero.}
\label{fig:timeDependent}
\end{figure}

Various time-dependent quantities are plotted in Fig.~\ref{fig:timeDependent}. 
``Energy Constraint'' refers to the right-hand side
of the constraint equation (\ref{eneInt}), and its deviation from $\Etot$ is a measure 
of the accuracy of the simulation. We see that the constraint deteriorates around
$t=40$ (as expected from the presence of large gradients), but before that 
the free energy $\Ftot=-d\Itot/dt$ reaches a nearly time-independent value of 
about $0.3$. 
The energy $E$ of the black hole at late times can be computed from the horizon 
radius using Eq.~(\ref{eneBH}) and, 
as seen in the figure, at late times is about $0.9$.

\begin{figure}[ht]
\includegraphics[scale=.75]{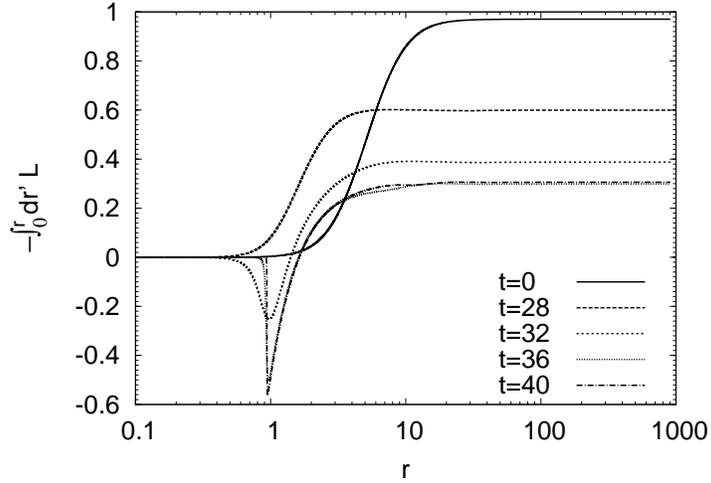}
\caption{Accumulation plots for the action (\ref{Itot}) at different moments of time.
(These show the negative of the Lagrangian contained in the ball of radius $r$, 
as a function of $r$.)}
\label{fig:Stot}
\end{figure}

Fig.~\ref{fig:Stot} shows where the free energy is accumulated. 
The growth from $-0.6$ to $0.3$ in the region outside the horizon
is the contribution
to $\Ftot$ from the static Schwarzschild exterior and, as expected, closely
matches the energy of the black hole. Note that while an outgoing wave with a crest
at $r\sim 10$ is clearly seen in Fig.~\ref{fig:f}, 
it practically does not contribute to
$\Ftot$. In other words, in this case, $\Ftot$ and $F$, the free energy of the black 
hole alone, nearly coincide. Combining the results of Figs.~\ref{fig:timeDependent} 
and \ref{fig:Stot}, we find that at late times $F\approx \frac{1}{3} E$.
Moreover, we see that the difference between
$E$ and $F$ comes from a thin shell near the horizon (the drop at $r\approx 1$ in
Fig.~\ref{fig:Stot}). 

\section{Conclusion}
The counting of powers of $\hbar$, described in the Introduction, shows that,
unlike the temperature or entropy, the free energy of a black hole remains finite 
in the classical limit and so may have
a purely classical interpretation. Our results indicate that this
interpretation is furnished by Eq.~(\ref{F}).
At present, our evidence for this is purely numerical and confined to a
specific type of black hole in a specific 5d theory. However, if the
correspondence turns out to be generic, it would be interesting to
see if it can be reproduced by an analytical argument.

We conclude with a remark on the quantum theory of the system described by the 
action (\ref{Itot}).
In spherical symmetry, quantization of gravity is straightforward: gravity has no
independent degrees of freedom and simply follows quantized matter. As we have
seen, in the Newtonian limit, gravitational collapse of an instanton is described
by the ordinary differential equation (\ref{eql}) for the modulus $\lambda$. 
(The Newtonian limit applies for $\lambda \gg r_0$.)
The corresponding Hamiltonian is that of a ``fall to the center'':
\be
H = \frac{\Pi^2}{2M} - \frac{4 \zeta M}{15 \lambda^2} \; ,
\label{H}
\ee
where $\Pi = M\dot{\lambda}$ is the momentum conjugate to $\lambda$, and
$M = 2 E_{\rm inst} = 16\pi^2 / \g^2$. The overall normalization of the Hamiltonian
can be restored by noting that, while the first nontrivial term in
the expansion (\ref{Psi}) gives $\Etot = E_{\rm inst}$, the second gives
a correction, which should equal the potential term in (\ref{H}).

As well known, a quantum problem described at large $\lambda$ by Eq.~(\ref{H})
is in one of the two different regimes, depending on the value of the coupling
\be
\sigma = \frac{8 \zeta M^2}{15 \hbar^2} \; .
\label{sigma}
\ee
For $\sigma > \frac{1}{4}$, there is an infinite number of bound states with small 
binding energies. For $\sigma \gg 1$, these states can form superpositions 
(wave packets) that move 
nearly classically. From the relation (\ref{sigma}), we see that this classical
limit corresponds to 
large values of $E_{\rm inst}$, while the critical 
$\sigma = \frac{1}{4}$ corresponds to $E_{\rm inst}$ being of the order of the 
5d Planck mass.

This work was supported in part by the U.S. Department of Energy through Grant
DE-FG02-91ER40681 (Task B).

\end{document}